\documentclass{aa}  

\usepackage{natbib,twoopt}
\usepackage[breaklinks=true, colorlinks=true, allcolors=blue]{hyperref} 
\makeatletter
\newcommandtwoopt{\citeads}[3][][]{\href{http://adsabs.harvard.edu/abs/#3}%
{\def\hyper@linkstart##1##2{}%
\let\hyper@linkend\@empty\citealp[#1][#2]{#3}}}
\newcommandtwoopt{\citepads}[3][][]{\href{http://adsabs.harvard.edu/abs/#3}%
{\def\hyper@linkstart##1##2{}%
\let\hyper@linkend\@empty\citep[#1][#2]{#3}}}
\newcommandtwoopt{\citetads}[3][][]{\href{http://adsabs.harvard.edu/abs/#3}%
{\def\hyper@linkstart##1##2{}%
\let\hyper@linkend\@empty\citet[#1][#2]{#3}}}
\newcommandtwoopt{\citeyearads}[3][][]%
{\href{http://adsabs.harvard.edu/abs/#3}
{\def\hyper@linkstart##1##2{}%
\let\hyper@linkend\@empty\citeyear[#1][#2]{#3}}}
\makeatother

\usepackage{geometry}

\usepackage{graphicx}
\usepackage{epstopdf}
\usepackage{rotating}
\usepackage{color}
\usepackage{fancybox}
\usepackage{pstricks}
\usepackage{pst-plot}
\usepackage{pstricks-add}
\usepackage{epsf}

\usepackage{graphicx} 
\usepackage{txfonts}
\usepackage{epsf} 
\begin{document}

    \title{X-ray obscuration from a variable ionized absorber in PG 1114+445}

   \author{R. Serafinelli
          \inst{1}\fnmsep\thanks{              \texttt{roberto.serafinelli@inaf.it}}
          \and
          V. Braito\inst{1,2}
          \and
          P. Severgnini\inst{1}
          \and
          F. Tombesi\inst{3,4,5,6}
          \and
          G. Giani\inst{1,7}
          \and
          E. Piconcelli\inst{6}
          \and
          R. Della Ceca\inst{1}
          \and
          \\ F. Vagnetti\inst{3,8}
          \and
          M. Gaspari\inst{9,10}
          \and
          F. G. Saturni\inst{6,11}
          \and
          R. Middei\inst{6,11}
          \and
          A. Tortosa\inst{12}
          }

   \institute{INAF - Osservatorio Astronomico di Brera, Via Brera 28, 20121, Milano, Italy \& Via Bianchi 46, Merate (LC), Italy
          \and
          Department of Physics, Institute for Astrophysics and Computational Sciences, The Catholic University of America, Washington, DC, 20064, USA
         \and
Dipartimento di Fisica, Universit\`a degli Studi di Roma ``Tor Vergata'', Via della Ricerca Scientifica 1, 00133, Roma, Italy
         \and
X-ray Astrophysics Laboratory, NASA/Goddard Space Flight Center, Greenbelt, MD 20771, USA
         \and
Department of Astronomy, University of Maryland, College Park, MD 20742, USA
        \and
INAF – Osservatorio Astronomico di Roma, Via Frascati 33, 00044 Monte Porzio Catone, Roma, Italy
         \and
Dipartimento di Fisica, Universit\`a degli Studi di Milano, Via Celoria, 16, 20133 Milano, Italy
         \and
INAF - Istituto di Astrofisica e Planetologia Spaziali, Via del Fosso del Cavaliere 100, 00133, Roma, Italy
        \and
INAF, Osservatorio di Astrofisica e Scienza dello Spazio, via P. Gobetti 93/3, 40129 Bologna, Italy
         \and
Department of Astrophysical Sciences, Princeton University, Princeton, NJ 08544, USA  
         \and
Space Science Data Center, Agenzia Spaziale Italiana, Via del Politecnico snc, 00133 Roma, Italy
         \and
N\'ucleo de Astronom\'ia de la Facultad de Ingenier\'ia, Universitad Diego Portales, Av. Ej\'ercito Libertador 441, Santiago, Chile
             }

   \date{Received XXX; accepted YYY}

 
  \abstract
   {Photoionized absorbers of outflowing gas are commonly found in the X-ray spectra of active galactic nuclei (AGN). While most of these absorbers are seldom significantly variable, some ionized obscurers have been increasingly found to substantially change their column density on a wide range of time scales. These $N_\text{H}$ variations are often considered as the signature of the clumpy nature of the absorbers. Here we present the analysis of a new {\it Neil Gehrels Swift Observatory} campaign of the type-1 quasar PG 1114+445, which was observed to investigate the time evolution of the multiphase outflowing absorbers previously detected in its spectra. The analyzed dataset consists of 22 observations, with a total exposure of $\sim90$ ks, spanning about $20$ months. During the whole campaign, we report an unusually low flux state with respect to all previous X-ray observations of this quasar. From the analysis of the stacked spectra we find a fully covering absorber with a column density $\log(N_\text{H}/\text{cm}^{-2})=22.9^{+0.3}_{-0.1}$. This is an order of magnitude higher than the column density measured in the previous observations. This is either due to a variation of the known absorbers, or by a new one, eclipsing the X-ray emitting source. We also find a ionization parameter of $\log(\xi/\text{erg cm s}^{-1})=1.4^{+0.6}_{-0.2}$. Assuming that the obscuration lasts for the whole duration of the campaign, i.e. more than $20$ months, we estimate the minimum distance of the ionized clump, which is located at $r\gtrsim0.5$ pc.}


   \maketitle
%

\section{Introduction}
\label{sec:intro}

\begin{table}
\centering
\begin{tabular}{lccccc}
\hline
Observation ID & Date & Exposure (s)\\
\hline
00011004001 & 2019-03-08 & 4909\\
00011004002 & 2019-03-08 & 4978\\
00011004003 & 2019-03-15 & 4258\\
00011004004 & 2019-03-15 & 4475\\
00011004005 & 2019-03-29 & 4249\\
00011004006 & 2019-03-30 & 4750\\
00011004007 & 2019-04-29 & 4957\\
00011004008 & 2019-04-29 & 4821\\
00011004009 & 2019-05-06 & 4313\\
00011004010 & 2019-05-06 & 1920\\
00011004011 & 2019-05-21 & 5152\\
00011004012 & 2019-05-21 & 4874\\
00011004013 & 2019-06-21 & 5070\\
00011004014 & 2019-06-21 & 5151\\
00011004015 & 2019-06-28 & 4404\\
00011004016 & 2019-06-28 & 4849\\
00011004017 & 2019-07-13 & 4051\\
00011004018 & 2019-07-13 & 4527\\
00089058001 & 2020-10-26 & 1804\\
00011004019 & 2020-12-06 & 1855\\
00011004020 & 2020-12-08 & 2453\\
00011004021 & 2020-12-08 & 2795\\
\hline
\end{tabular}
\caption{The list of X-ray observations taken by {\it Swift}-XRT, with OBSID, observation date and exposure.}
\label{tab:data}
\end{table}

Active galactic nuclei (AGN) are extremely luminous extragalactic objects, located at the center of their host galaxies, and powered by the accretion of matter onto a supermassive black hole (SMBH). AGN are now considered a major player in shaping their host galaxy during its evolution, since many of its properties are correlated with the mass of the central SMBH \citep[e.g.,][]{ferrarese00,haring04,gaspari19}. Outflows are one of the main mechanisms by which the black hole is believed to transport its energy to large distances \citep[e.g.,][]{king15,fiore17,cicone18,laha21}. Such winds are commonly found in AGN spectra, at many wavelengths \citep[e.g.,][]{gibson09,harrison14,cicone14,vietri18}.\\
\indent In the X-ray band, absorption features are the typical signature of the presence of outflows. Low-ionization absorbers ($\log(\xi/\text{erg cm s}^{-1})\lesssim2$), characterized by a low outflow velocity ($v_\text{out}\sim100-1000$ km s$^{-1}$), are found in about $65\%$ of soft X-ray ($E\lesssim2$ keV) spectra of nearby AGN, and they are often known as warm absorbers \citep[WAs, e.g.,][]{halpern84,blustin05,mckernan07,laha14}. More than $30\%$ X-ray-detected AGN show evidence of the presence of ultra-fast outflows \citep[UFOs, e.g.,][]{chartas02,pounds03a,pounds03b,braito07,tombesi10, gofford13, tombesi15,nardini15,ballo15}. UFOs are extremely ionized ($\log(\xi/\text{erg cm s}^{-1})\sim3-6$) absorbers, detected as blueshifted absorption lines of Fe {\footnotesize XXV} and {\footnotesize XXVI}, with typical outflow velocity of $v_\text{out}\sim0.1c$, but capable of reaching near-relativistic values of $\sim0.5c$ \citep[e.g.,][]{reeves18,luminari21}.\\
\indent While UFOs are extremely variable \citep[e.g.,][]{matzeu17}, variability of soft X-ray ionized absorbers is found less frequently. In some cases, some sources are found in a state with diminished X-ray flux, due to an increase of the column density of the obscuring medium. The absorbers are sometimes found to be persistent for about a decade \citep[e.g., NGC 5548,][]{kaastra14}. In other cases, the obscurer has a much shorter variability timescale \citep[e.g.,][]{severgnini15,matzeu16,mehdipour17,middei20}. In a few other cases, a source was found in an higher flux state than its usual one, due to a diminished obscuring power of the WA \citep[e.g.,][]{braito14}. All these cases suggest a clumpy structure for the ionized and possibly outflowing absorbers around the AGN, which is predicted by duty-cycle theoretical models such as chaotic cold accretion \citep[CCA, e.g.,][]{gaspari13,gaspari17}. Such clumpy material can be part of the multiphase rain condensing out of the hot halo, which then eventually contributes to the feeding component, alongside the feedback channel. Indeed, both feeding and feedback processes are expected to be tightly self-regulated over the cosmic time and over nine orders of magnitude in spatial scale \citep[see e.g.,][for a review]{gaspari20}. \\
\indent The type-1 quasar PG 1114+445 \citep[$z=0.144$,][]{hewett10} has an estimated black hole mass of $\log(M/M_\odot)\simeq8.8$ and bolometric luminosity of $\log(L_\text{bol}/\text{erg s}^{-1})\simeq45.7$ \citep{shen11}. Hence, the source is accreting with an Eddington ratio of $\log(L_\text{bol}/L_\text{Edd})\simeq-1.14$. The source is well known to host multiple absorbers, in both the UV and X-ray bands. An UV observation taken in 1996 with the Faint Object Spectrograph (FOS) on board of the Hubble Space Telescope (HST) was able to detect Ly$\alpha$ and C {\footnotesize IV} absorption lines in the spectrum of this quasar. For such lines, an outflowing velocity of $\sim530$ km s$^{-1}$ was measured \citep{mathur98}. The observation was simultaneous with an Advance Satellite for Cosmology and Astrophysics (ASCA) pointing, that highlighted the presence of an ionized WA, and showed marginal evidence of the presence of an absorption line at $E\sim7.3$ keV \citep{george97}. The UV absorption lines and the X-ray warm absorber have similar ionizations and therefore they likely trace the same material \citep{mathur98}.\\
\indent The source was observed again in 2002 with {\it XMM-Newton} \citep{jansen01}, revealing that the absorption complex consisted of two WA layers \citep{ashton04,piconcelli05}. A further {\it XMM-Newton} campaign of 11 observations was performed in 2010. The data from this campaign, together with a re-analysis of the 2002 observation \citep[][hereafter Paper I]{serafinelli19}, found that one of the two absorbers has typical WA parameters -- i.e. column density $N_\text{H}\simeq7.6\times10^{21}$ cm$^{-2}$, and ionization parameter $\log(\xi/\text{erg cm s}^{-1})\simeq0.35$ with velocity below the energy resolution -- likely associated with the UV absorber \citep{mathur98}. The second absorber shares very similar parameters with the WA ($N_\text{H}\sim3\times10^{21}$ cm$^{-2}$ and $\log(\xi/\text{erg cm s}^{-1})~\sim0.5$), with the exception of the outflow velocity, which is high enough to be detected with the EPIC cameras on board {\it XMM-Newton} ($v_\text{out}/c=0.12\pm0.03$). Low-ionization fast outflows were also found in other sources \citep[e.g.,][]{longinotti15,pounds16,reeves20}. In addition, a high-ionization UFO ($v_\text{out}/c=0.15\pm0.04$, consistent with the fast absorber) was found in three spectra. This evidence led to the interpretation of a UFO pushing and entraining the host galaxy interstellar medium to a comparable velocity, producing a so-called entrained ultra-fast outflow (E-UFO, Paper I).\\
\indent In this paper we analyze the data taken during a recent {\it Neil Gehrels Swift Observatory} (hereafter {\it Swift}) X-Ray Telescope \citep[XRT,][]{gehrels04} campaign, which was proposed to study the possible variability of the absorbers found in Paper I, on timescales from days to months. In Sect.~\ref{sec:data} we describe how the data was prepared for the analysis. In Sect.~\ref{sec:var} we analyze the X-ray amplitude and spectral variability, while in Sect.~\ref{sec:spec} we present a detailed spectroscopy of the source. In Sect.~\ref{sec:uvar} we analyze the variability of the X-ray/UV ratio. We summarize and discuss our results in Sect.~\ref{sec:discussion}.\\
\indent Throughout the paper, we adopt the following cosmology: $\Omega_m=0.3$, $\Omega_\Lambda=0.7$, and $H_0=70$ km s$^{-1}$ Mpc$^{-1}$. All uncertainties are reported at a $90\%$ confidence level.

\section{Observations and data reduction}
\label{sec:data}

\begin{figure}
\centering
\includegraphics[scale=0.45]{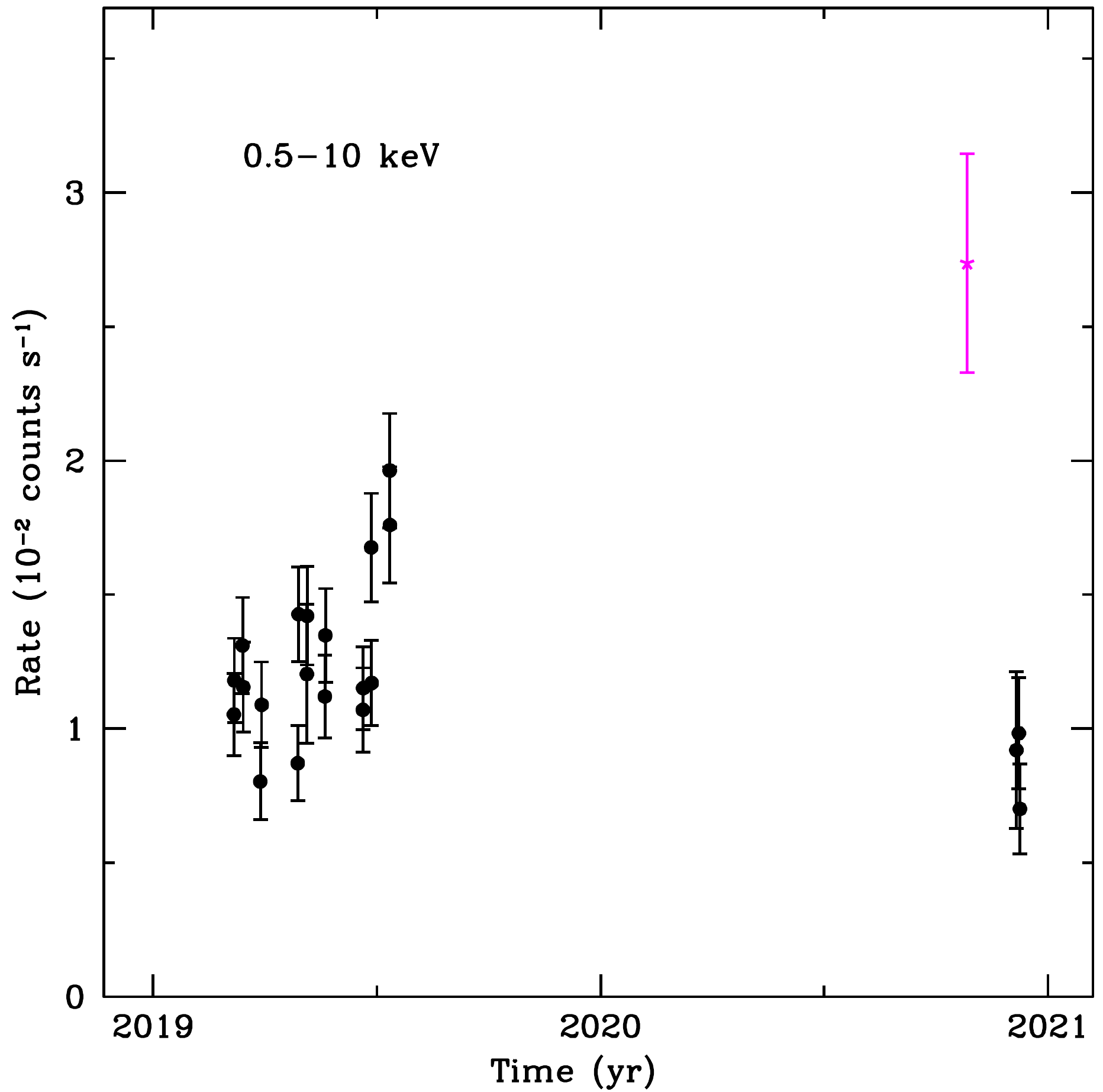}
\caption{{\it Swift}-XRT count rates in the energy band $E=0.5-10$ keV. The magenta star represents OBSID 00089058001, that was not part of our campaign.}
\label{fig:cr}
\end{figure}

\begin{figure*}
\centering
\includegraphics[scale=0.45]{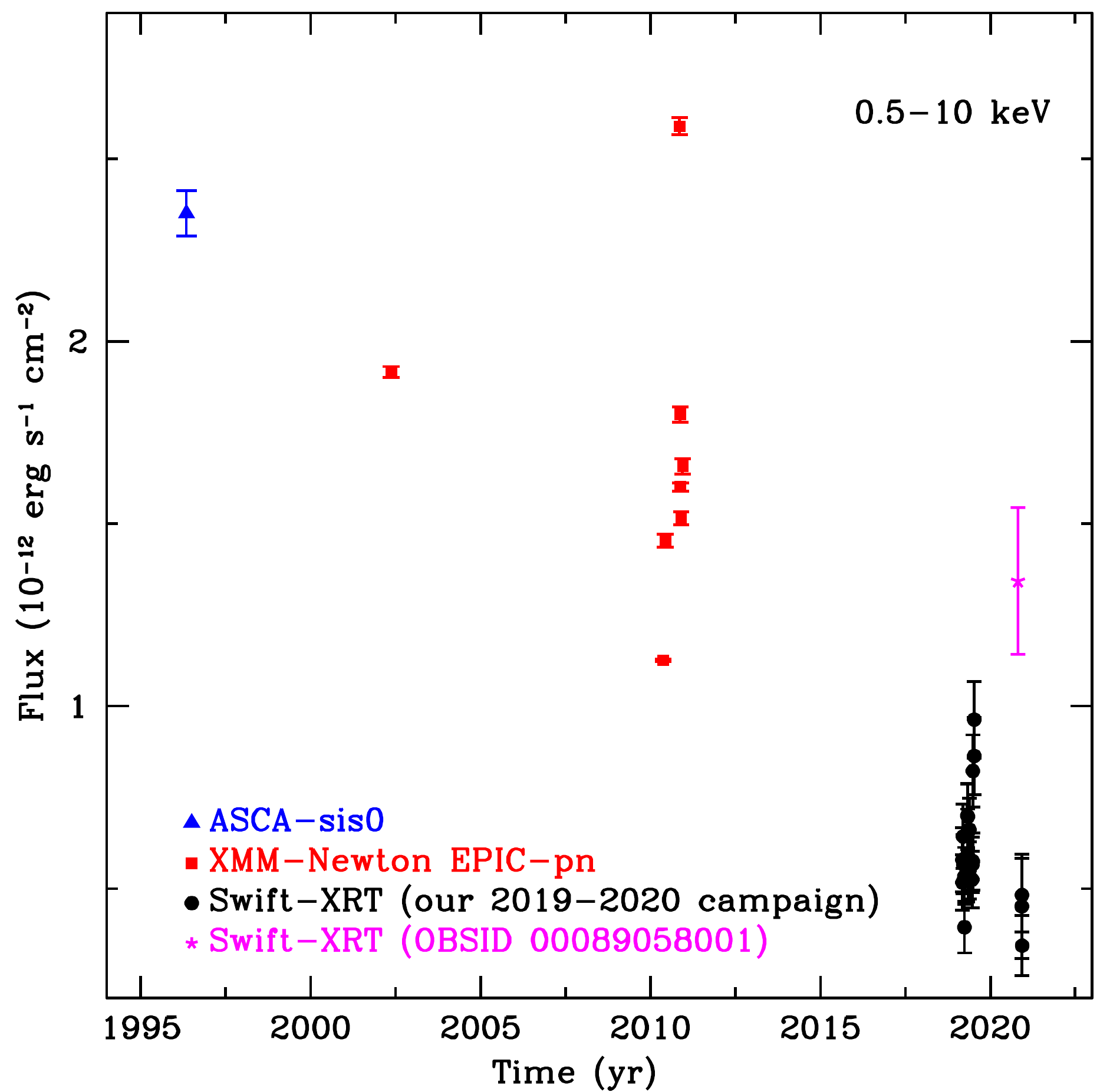}\includegraphics[scale=0.447]{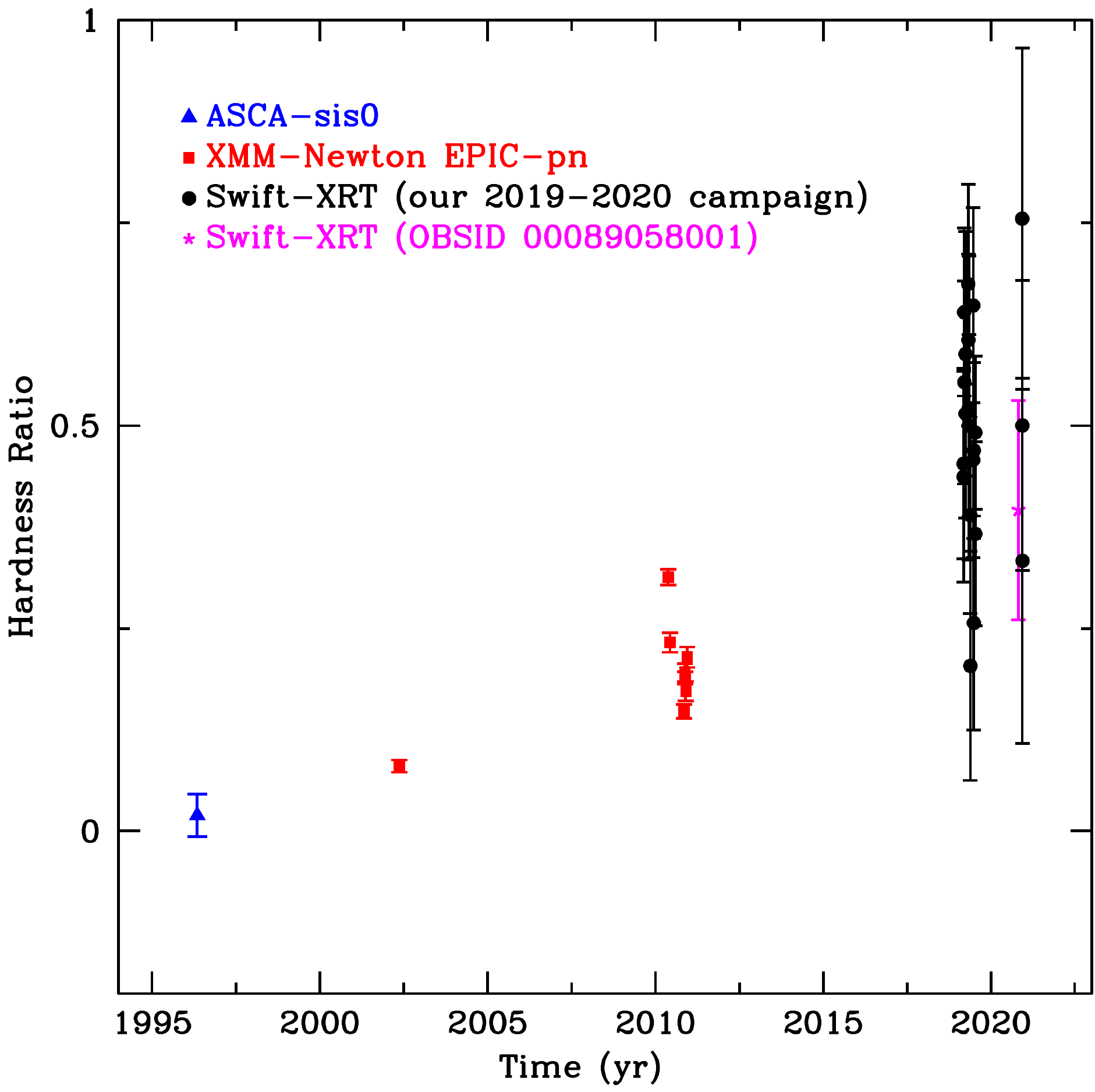}
\caption{Left panel. Full-band ($E=0.5-10$ keV) light curve, using ASCA data (blue triangle), {\it XMM-Newton} (red squares), our {\it Swift}-XRT campaign (black circles) and additional archival {\it Swift}-XRT observation (magenta star). Right panel. The time variability of the hardness ratio HR=$(H-S)/(H+S)$, where $S$ and $H$ are the $0.5-2$ and $2-10$ keV count rates, respectively, is shown. There is a noticeable hardening of the spectrum in the most recent data. The count rates of the ASCA and EPIC-pn observations were converted into {\it Swift}-XRT count rates following the procedure described in Sect.~\ref{sec:var}. }
\label{fig:lcfull}
\end{figure*}

\indent The observations were performed during {\it Swift} Cycle 15 (PI: Serafinelli) from March to July $2019$ (OBSID 00011004001 to 00011004018), and then three additional observations were taken $15$ months later, in December $2020$ (OBSID 00011004019 to 00011004021). The observations were spaced by $7$, $15$ and $30$ days, in order to analyze possible short-time variability within the campaign. A further archival observation, OBSID 00089058001, not part of our campaign, taken in October 26th 2020, is also considered here. Most observations are about $\sim4-5$ ks long, with some observations being only $\sim2$ ks long. The list of {\it Swift}-XRT observations is shown in Table~\ref{tab:data}. The total exposure time is $\sim90$ ks.\\
\indent For each {\it Swift}-XRT observation, the source and background spectra were extracted using the {\footnotesize HEASOFT} task {\footnotesize XSELECT}. The source spectrum was extracted from a circular area of $40"$ radius around the object, while the background spectrum was extracted from two source-free circular areas of $40"$ radius each in the proximity of the source. Ancillary files were produced using the {\footnotesize XRTMKARF} task, while the response was taken from the {\footnotesize HEASOFT CALDB} repository.\\
\indent Each {\it Swift} observation provided a UltraViolet and Optical Telescope (UVOT) pointing, with a single filter, centered on the source. The source monochromatic flux was measured within circular apertures of $5"$ radius, while the background was extracted from  an annulus region, centered on the source, with internal radius of $15"$ and external radius of $40"$, using the {\footnotesize UVOTSOURCE} task.

\section{X-ray variability}
\label{sec:var}

The {\it Swift}-XRT count rate light curve of PG 1114+445 in the $0.5-10$ keV full band is shown in Fig.~\ref{fig:cr}. The average count rate is $\sim1.5\times10^{-2}$ cts s$^{-1}$. We compare the current data set with archival X-ray pointings from ASCA, taken in 1996 and {\it XMM-Newton}, taken in 2002 and 2010. The products of the ASCA observation were downloaded from the Tartarus database\footnote{\tt \url{https://heasarc.gsfc.nasa.gov/FTP/asca/data/tartarus/products/74072000/74072000_gsfc.html}} \citep{turner01}, while details on the {\it XMM-Newton} data reduction can be found in Paper I. In order to compare data taken with different telescopes, we convert the X-ray count rates into fluxes. All fluxes are obtained with the web tool {\footnotesize WebPIMMS}\footnote{\tt \url{https://heasarc.gsfc.nasa.gov/cgi-bin/Tools/w3pimms/w3pimms.pl}}, adopting a simple AGN X-ray spectrum, composed by a typical $\Gamma=1.9$ powerlaw \citep[e.g.,][]{corral11,serafinelli17}, at the redshift of the source ($z=0.144$), with Galactic absorption \citep[$N_\text{H,Gal}=1.87\times10^{20}$ cm$^{-2}$, ][]{hi4pi16}. The full band fluxes as a function of the observation time are plotted in the left panel of Fig.~\ref{fig:lcfull}, where an evident decrease of the X-ray flux can be already identified with a visual inspection.\\
\indent The possible presence of obscuration can be investigated by the use of the hardness ratio (HR), which we define as $(H-S)/(H+S)$, where $S$ is the count rate in the soft band ($E=0.5-2$ keV), and $H$ the count rate in the hard band ($E=2-10$ keV). In order to deal with comparable HR, which are typically different due to the different response of the instruments, we convert all the count rates to the ones of {\it Swift}-XRT, using {\footnotesize WebPIMMS} and assuming again a simple absorbed $\Gamma=1.9$ powerlaw at $z=0.144$, with $N_\text{H,Gal}=1.87\times10^{20}$ cm$^{-2}$. As shown in Fig.~\ref{fig:lcfull} (right panel), there is an average hardening of the spectrum in the {\it Swift}-XRT observations, whereas the HR values during the whole {\it Swift}-XRT campaign are roughly consistent within the error bars. During the observation performed on 26th October 2020, i.e. the magenta point in Figs.~\ref{fig:cr} and \ref{fig:lcfull} the flux was higher than during the rest of the 2019-2020 campaign. However, the HR is still consistent with the value of the other observations of the {\it Swift} campaign. The increased HR with respect to the {\it XMM-Newton} observations suggests that the lower flux state is not due to a change in the primary continuum, but likely to a variation of the column density of one or more of the known absorbers found in Paper I. 

\section{X-ray spectral analysis}
\label{sec:spec}

\begin{figure}
\centering
\includegraphics[scale=0.36]{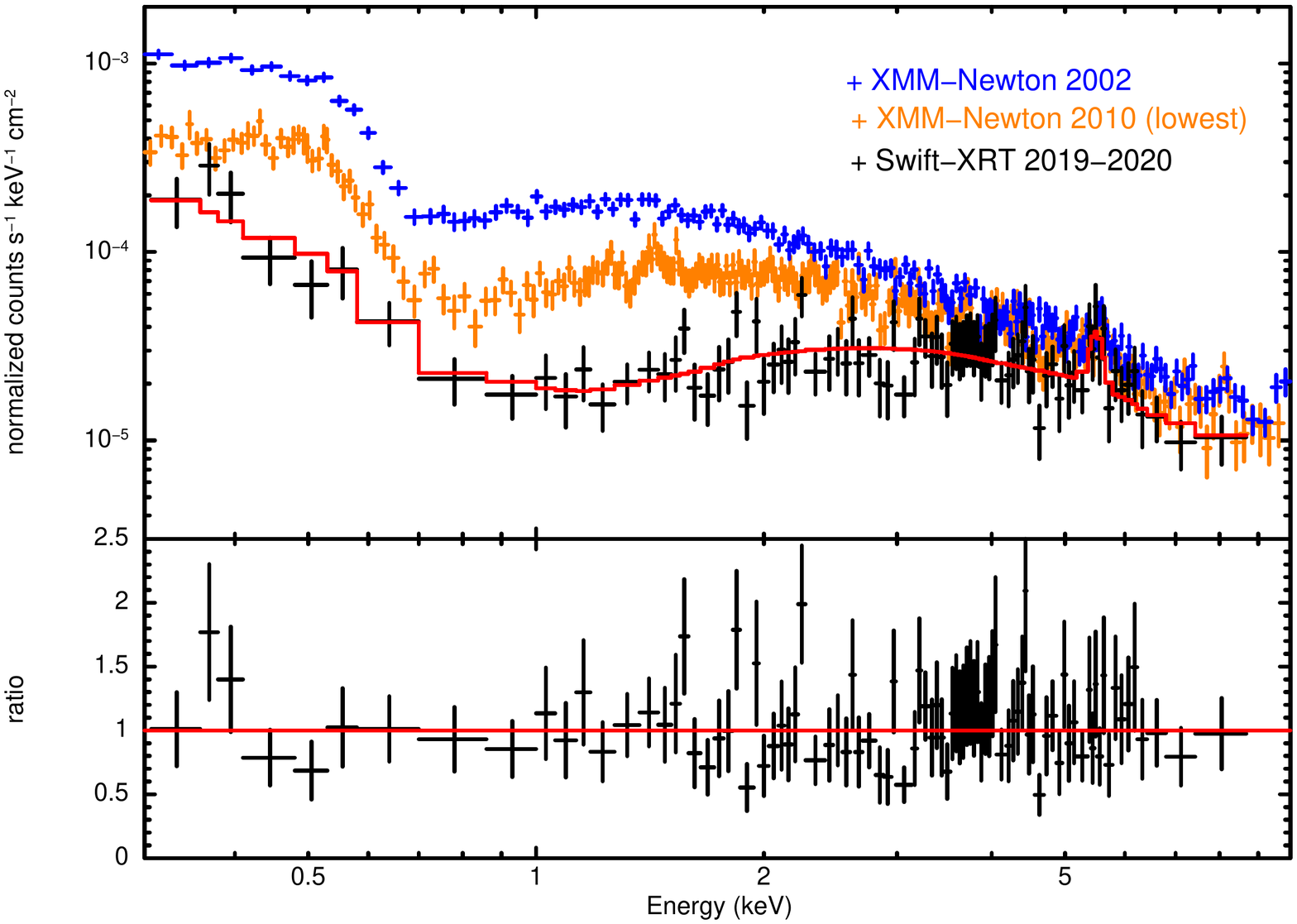}
\caption{Composite {\it Swift}-XRT spectrum (black), with its best-fit model (upper panel) and data-to-model ratios (lower panel). For comparison, we also show the 2002 EPIC-pn observation (blue) and the lowest-flux spectrum of the 2010 campaign (orange).}
\label{fig:spec}
\end{figure}

\begin{figure}
\centering
\includegraphics[scale=0.37]{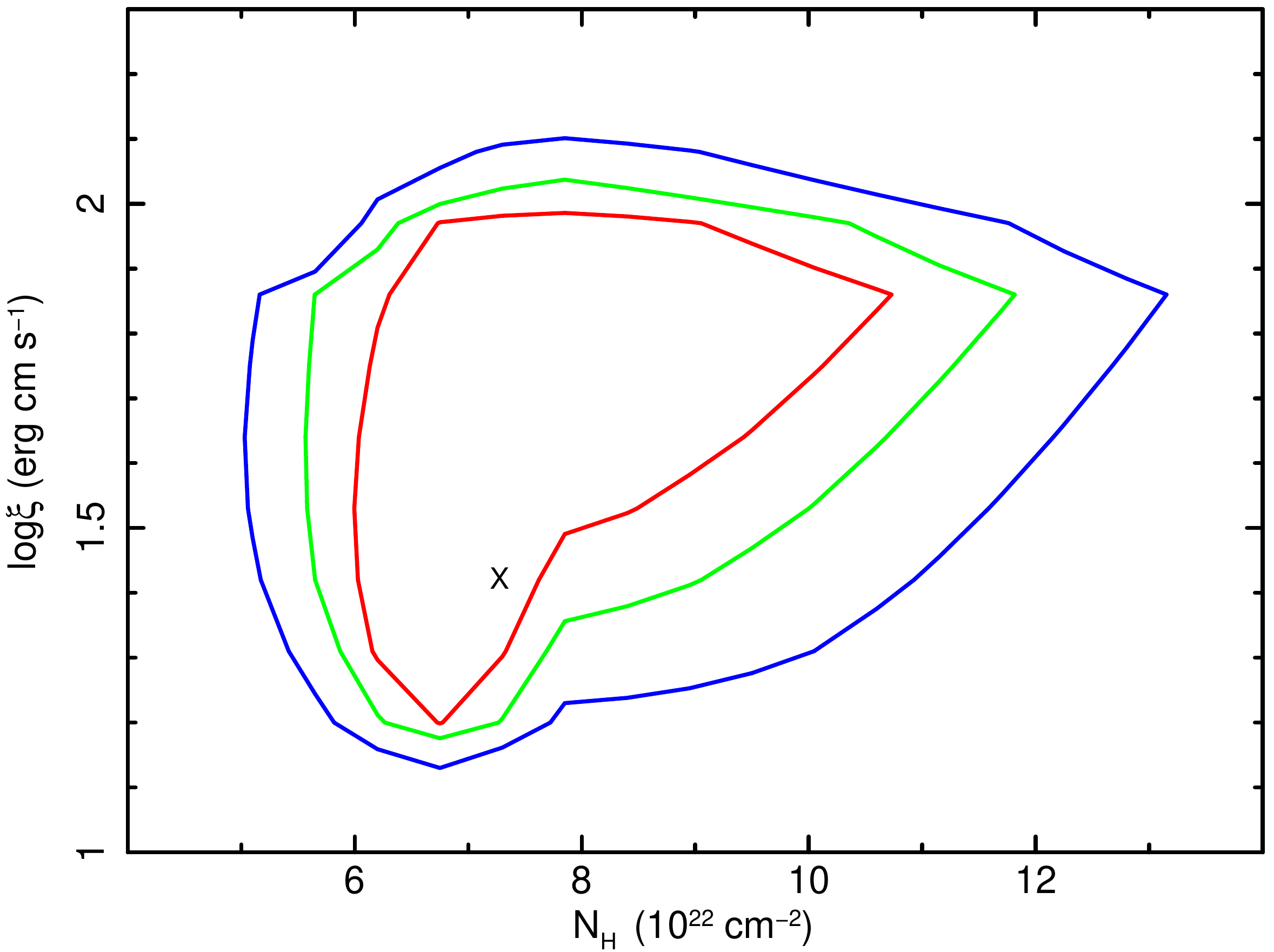}
\caption{Contour plot for the ionization $\log\xi$ and the column density $N_\text{H}$. The red line represents the $1\sigma$ confidence level ($68\%$), while the green and blue lines represent the $2\sigma$ ($95\%$) and $3\sigma$ ($99.7\%$) confidence levels. The X marks the best-fit values.}
\label{fig:contnhxi}
\end{figure}

Given that all {\it Swift}-XRT observations have small number of photon counts, in order to maximize the signal-to-noise ratio we combine all the observations into a single spectrum using the {\footnotesize HEASOFT} task {\footnotesize ADDSPEC}. We bin the combined spectrum, which consists of $\sim1300$ net counts, to have at least $10$ counts per energy channel.\\
\indent All spectral fits are performed using {\footnotesize XSPEC v12.10} \citep{arnaud96}, adopting a C-statistic, given the low number of photons for each bin \citep[e.g.,][]{kaastra17}. We first attempt to fit the combined spectrum with a simple model, consisting in a power-law component and  and an emission line, with Galactic absorption characterized by a fixed column density $N_\text{H,Gal}=1.87\times10^{20}$ cm$^{-2}$ \citep[][]{hi4pi16}: {\tt Tbabs $\times$ (powerlaw $+$ zgauss)}. The fit results in an unacceptable C-statistic: $C/\text{dof}=229/106$, where dof is the degree of freedom, and an unrealistic photon index, $\Gamma=0.4\pm0.1$.\\
\indent Based on previous analyses of PG 1114+445 \citep[Paper I]{george97,ashton04,piconcelli05}, we include an ionized absorber, using a partial covering model, {\footnotesize ZXIPCF} \citep{reeves08}. The assumed model is therefore {\tt Tbabs $\times$ zxipcf $\times$ (powerlaw $+$ zgauss)}. The addition of this ionized absorption component significantly improves the goodness of fit, with a C-stat of $C/\text{dof}=106/103$. The best-fit value found for the photon index is $\Gamma=1.6^{+0.3}_{-0.4}$, which is consistent with the values measured in Paper I. We find that the absorber is moderately ionized, $\log(\xi/\text{erg cm s}^{-1})=1.4_{-0.2}^{+0.6}$, and almost fully covering, with covering factor $C_f=0.96^{+0.03}_{-0.07}$), with a column density of $N_\text{H}=7_{-1}^{+5}\times10^{22}$ cm$^{-2}$. Such value is a factor of $\sim10$ larger than the column density of any low-ionization absorber detected in Paper I, confirming that the flux decrease is mainly due to a column density increase with respect to the $2010$ observations. The spectrum, the best-fit model and the data-to-model residuals are shown in Fig.~\ref{fig:spec}. The confidence contour plot of the joint errors of the ionization parameter vs $N_\text{H}$ is shown in Fig.~\ref{fig:contnhxi}. In particular, the column density is larger than the $N_\text{H}$ of any soft X-ray absorber detected in Paper I at $3\sigma$ confidence level.\\

\begin{table}[h]
\centering
\begin{tabular}{lcc}
\hline
\\
Parameter & Value\\
\\
\hline
\\
$\log N_\text{H}$ (cm$^{-2}$) & $22.9_{-0.1}^{+0.3}$\\
\\
$\log\xi$ (erg cm s$^{-1}$) &  $1.4_{-0.2}^{+0.6}$\\
\\
$C_f$ & $0.96_{-0.07}^{+0.03}$\\
\\
$\Gamma$ & $1.6^{+0.3}_{-0.4}$ \\
\\
norm (cts s$^{-1}$) & $3.1_{-1.4}^{+2.2}\times10^{-4}$\\
\\
$E_{\text{K}\alpha}$ (keV) & $6.28\pm0.08$\\
\\
norm$_{\text{K}\alpha}$ (cts s$^{-1}$) & $7_{-3}^{+4}\times10^{-6}$\\
\\
Cstat/dof & $106/103$\\
\\
\hline
\end{tabular}
\caption{Summary of best-fit parameters for our best model: {\tt TBabs $\times$ zxipcf $\times$ (powerlaw $+$ zgauss)}. Errors are given at $90\%$ confidence level.}
\label{tab:bestfit}
\end{table}

\section{UV variability}
\label{sec:uvar}

\begin{table}
\centering
\begin{tabular}{lcc}
\hline
\\
OBSID & UV Filter & Flux ($10^{-15}$ erg cm$^{-2}$ s$^{-1}$ $\AA^{-1}$)\\
\\
\hline
00011004001 & UVM2 & $4.19\pm0.07$\\
00011004002 & UVW2 & $5.01\pm0.07$\\
00011004003 & UVM2 & $4.39\pm0.07$\\
00011004004 & UVW2 & $4.82\pm0.07$\\
00011004005 & UVM2 & $4.38\pm0.07$\\
00011004006 & UVW2 & $5.18\pm0.08$\\
00011004007 & UVM2 & $3.92\pm0.06$\\
00011004008 & UVW2 & $4.58\pm0.07$\\
00011004009 & UVM2 & $3.92\pm0.06$\\
00011004010 & UVW2 & $4.42\pm0.08$\\
00011004011 & UVM2 & $3.91\pm0.06$\\
00011004012 & UVW2 & $4.39\pm0.07$\\
00011004013 & UVM2 & $4.06\pm0.07$\\
00011004014 & UVW2 & $4.95\pm0.07$\\
00011004015 & UVM2 & $4.35\pm0.07$\\
00011004016 & UVW2 & $5.13\pm0.08$\\
00011004017 & UVM2 & $4.17\pm0.07$\\
00011004018 & UVW2 & $4.96\pm0.07$\\
00089058001 & U & $3.68\pm0.06$\\
00011004019 & UVW2 & $5.78\pm0.10$\\
00011004020 & UVW1 & $4.67\pm0.08$\\
00011004021 & U & $3.83\pm0.06$\\
\hline
\end{tabular}
\caption{List of flux densities in units of $10^{-15}$ erg cm$^{-2}$ s$^{-1}$ $\AA^{-1}$ for each {\it Swift}-UVOT pointing. The corresponding UVOT filter is also reported.}
\label{tab:uvot}
\end{table} 

\begin{figure}
\includegraphics[scale=0.46]{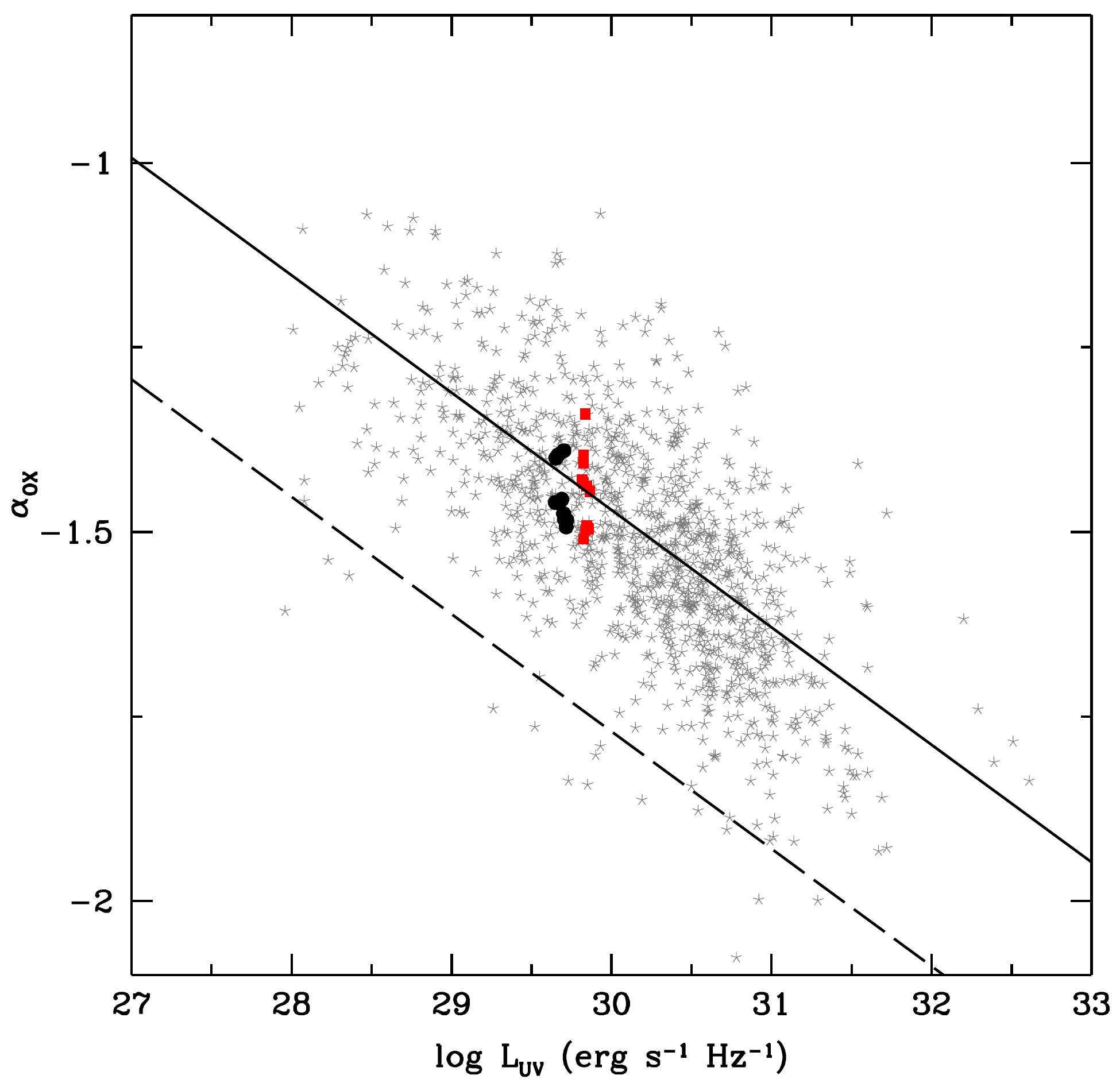}
\caption{Plot of the $\alpha_\text{ox}$ index versus the UV luminosity, using unabsorbed X-ray luminosities. The grey points are the quasars analyzed in \citet{chiaraluce18}, with the linear best-fit shown as a solid line. The dashed line is the X-ray weak limit defined by \cite{pu20}. The red squares correspond to the {\it XMM-Newton} observations, while the black points correspond to those performed by {\it Swift}.}
\label{fig:alox}
\end{figure}

The UV observations are provided by the {\it XMM-Newton} Optical Monitor (OM) and by {\it Swift}-UVOT. The OM data were retrieved from the XMM Serendipitous Ultraviolet Source Survey \citep[XMM-SUSS,][]{page12}. All UVOT filters do not show evidence of relevant variability between 2010 and 2019-2020 (see Table~\ref{tab:uvot}).\\
\indent We computed the X-ray/UV ratio $\alpha_\text{ox}$ \citep[e.g.,][]{vignali03,vagnetti10,lusso16,chiaraluce18}, in order to analyze the relative variability of the X-ray and UV bands, which is defined as $$\alpha_\text{ox}=\frac{\log L(2\text{ keV})-\log L (2500\,\AA)}{\log\nu (2\text{ keV})-\log\nu (2500\,\AA)},$$ where $\nu(2$ keV$)$ and $\nu(2500\,\AA)$ are the frequencies (in Hz) in the rest frame, corresponding to an energy of $2$ keV and a wavelength of $2500\,\AA$, respectively. The $2$ keV luminosity of the XRT observations was computed using {\footnotesize XSPEC}. We adopted the best-fit model (see Sect.~\ref{sec:spec}) and froze every parameter with the exception of the column density $N_\text{H}$ of the absorber and the normalization of the power law. We fit each snapshot with this model to obtain a best-fit model as accurate as possible, and then we remove the absorption component and use the {\footnotesize LUMIN} task between $1.99$ and $2.01$ keV, from which we derive the intrinsic X-ray luminosity $L(2$ keV$)$. Since the UVOT observations were performed with a single source-centered filter, we compute the luminosity at $2500\,\AA$ interpolating a spectral energy distribution obtained averaging thousands of Sloan Digital Sky Survey (SDSS) observations \citep{richards06}. We conservatively compute $L (2500\,\AA)$ only for those observations performed with the UVM2 filter, which is centered at $\sim2250$ $\AA$, corresponding to a rest-frame wavelength of $\sim2510$ $\AA$, and therefore close enough to $2500$ $\AA$. The fluxes are finally corrected for reddening due to the Galactic extintion adopting $E(B-V)=0.0264$ \citep{guver09}.\\
\indent The result is shown in Fig.~\ref{fig:alox}. The {\it Swift} observations lie on the ensemble trend found by \citet{chiaraluce18}, which is represented by a solid line. We note that none of the points lie on the X-ray weak limit (dashed line), which is set $0.3$ dex below the best-fit line \citep{pu20}. This is a further proof that the low flux state of the source is clearly driven by the obscuration and not by an intrinsic luminosity change.

\section{Summary and discussion}
\label{sec:discussion}

\begin{figure}
\centering
\includegraphics[scale=0.36]{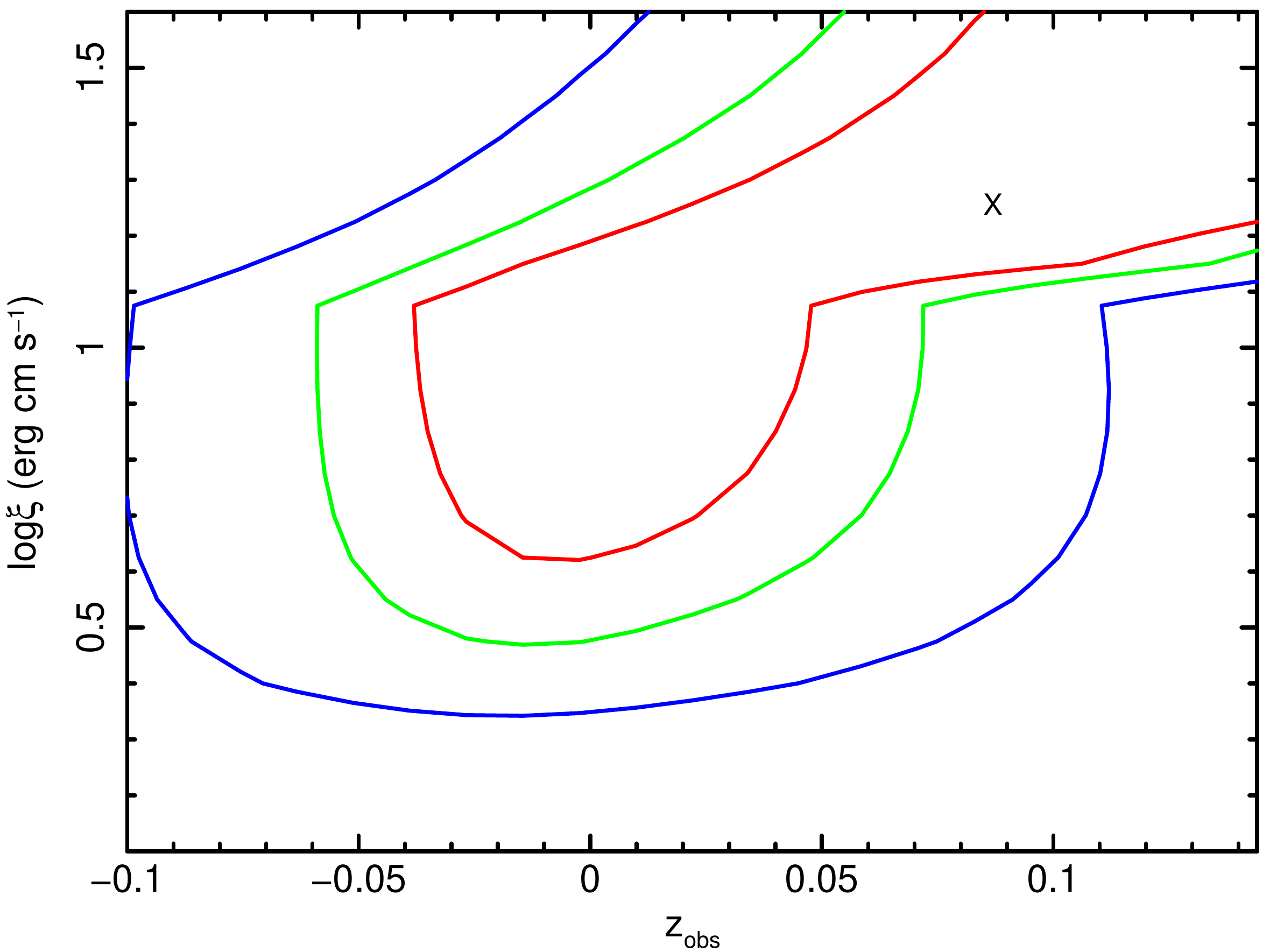}
\caption{Contour plots between the redshift of the absorber in the observer frame $z_\text{obs}$ and the ionization parameter $\log\xi$. The red, green and blue line represent the $1\sigma$ ($68\%$), $2\sigma$ ($95\%$) and $3\sigma$ ($99.7\%$) confidence levels, respectively.}
\label{fig:cont}
\end{figure}

\begin{figure}
\centering
\includegraphics[scale=0.44]{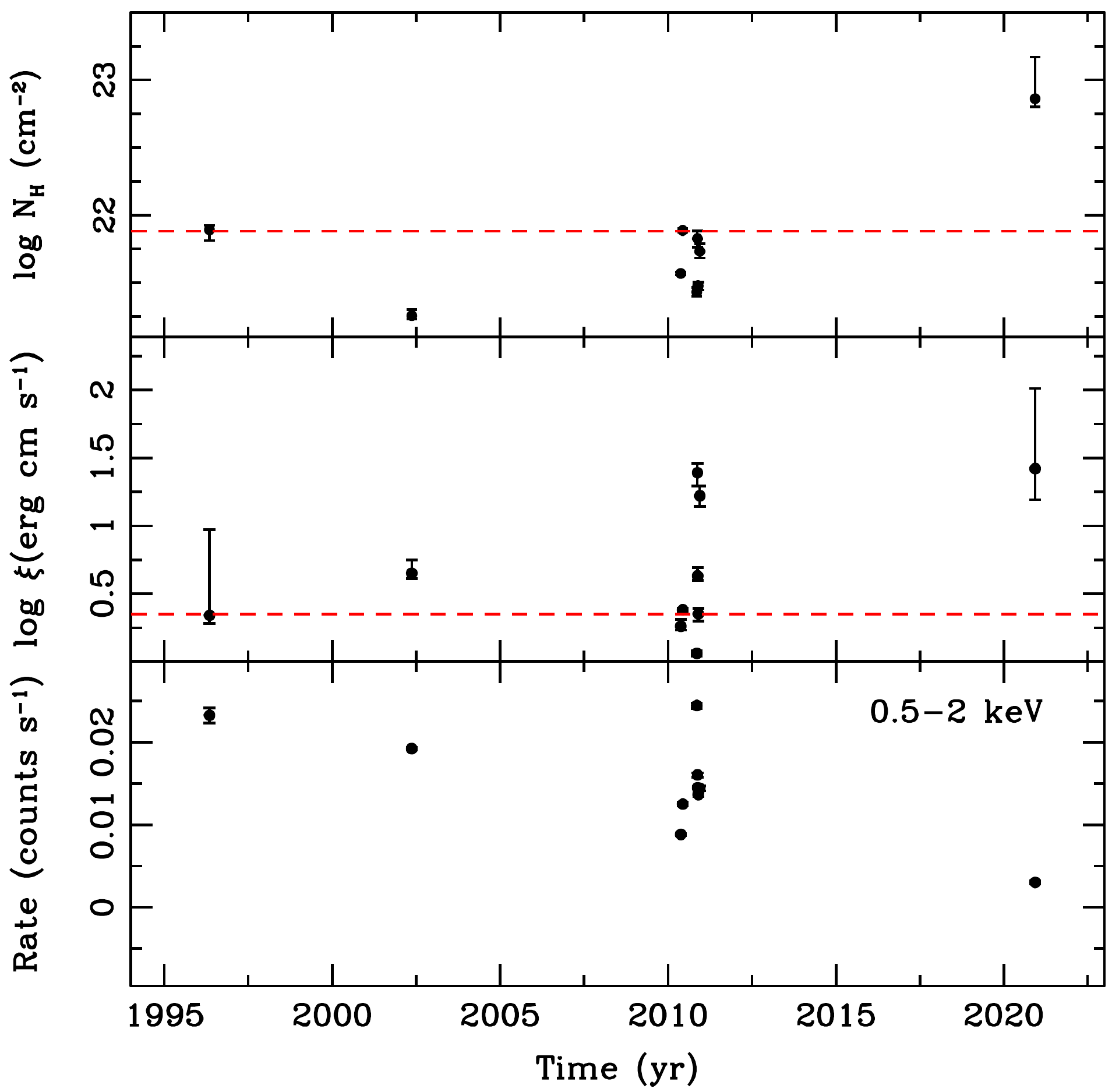}
\caption{Time variability of the absorber parameters $N_\text{H}$ (top) and $\xi$ (middle). For all observations prior to 2019, the black points are referred to the variable E-UFO, while the constant WA is represented by a red dashed line. In the lower panel, the light curve of soft X-ray ($E=0.5-2$ keV) count rates is shown, where the XMM rates were converted into {\it Swift}-XRT count rates using {\footnotesize WebPIMMS}.}
\label{fig:sum}
\end{figure}

In this work, we have presented the analysis of the {\it Swift} observations of the type-1 quasar PG 1114+445, performed $10$ years after the {\it XMM-Newton} observations analyzed in Paper I. The source is found in a strongly reduced flux state (Fig.~\ref{fig:lcfull}), which can be ascribed to an absorption increase by an obscuring material. Indeed, the spectral analysis highlights the presence of ionized ($\log\xi/\text{erg cm s}^{-1}\sim1.4$) material, characterized by a column density $N_\text{H}\sim7\times10^{22}$ cm$^{-2}$. No substantial variations on the unabsorbed X-ray luminosity of the source is found.\\
\indent In Paper I two absorbers were found in the soft X-rays, a slow and constant WA with $N_\text{H}\sim7\times10^{21}$ cm$^{-2}$, and a mildly-relativistic absorber, identified with an entrained ultra-fast outflow (E-UFO), with variable column density with median value $N_\text{H}\sim3\times10^{21}$ cm$^{-2}$ and a dispersion of $\sim6\times10^{20}$ cm$^{-2}$. Both these column densites are a factor of $10$ smaller than the one found in this work. If we allow for an outflowing velocity of this ionized absorber, we obtain that the best-fit value of the redshift of the absorber in the observer frame is $z_\text{obs}\sim0.09$, which corresponds to $v_\text{out}\sim0.05c$. However, the low statistic of the data set does not allow to successfully measure both the ionization parameter and the velocity, which are notoriously affected by degeneracy, as shown by their contour plot in Fig.~\ref{fig:cont}. We note that lower values of the absorber redshift are allowed at $1\sigma$ confidence level, down to $z_\text{obs}\simeq-0.03$ ($v_\text{out}\simeq0.16c$), consistent with both the E-UFO and the WA measured in Paper I. In addition, the ionization parameter can be as low as $\log\xi\lesssim0.4$ at $3\sigma$ confidence level, consistent with the E-UFO median value of 2010. This suggests that {\it Swift}-XRT is not able to resolve the complex structure of the absorbers measured in Paper I, but it may be hidden by the dominant absorber that we observe in this campaign.\\
\indent We can estimate a lower limit for its distance using the equation by \cite{risaliti02}: 

\begin{equation}
r\simeq4\times10^{16} \frac{M_\text{BH}}{10^M_\odot}\left(\frac{n}{10^9\text{ cm}^{-3}}\frac{t}{1\text{ day}}\frac{N_\text{H}}{5\times10^{22}\text{ cm}^{-2}}\right)^2\text{cm}\\
\label{eq:dist}
\end{equation}
$$\simeq2\times10^{23}n_9^2\;\text{cm,}$$ where $M_\text{BH}\simeq6.3\times10^8M_\odot$ is the black hole mass, $n$ is the density of the clump ($n_9$ in units of $10^9$ cm$^{-3}$), $N_\text{H}\simeq7\times10^{22}$ cm$^{-2}$ the column density of the clump, and $t$ is the duration of the obscuration. We assume here that the obscuration lasts during the whole campaing, including the times between 2019 and 2020 when the source was not observed. Therefore, we adopt $t\gtrsim560$ days (rest-frame) as the duration of the obscuration, which is only a lower limit since it could last beyond the end of the campaign. The density of the clump can be replaced by inverting the definition of ionization parameter: 
\begin{equation}
n=\frac{L_\text{ion}}{r^2\xi},
\label{eq:densy}
\end{equation}
where $L_\text{ion}\simeq1.5\times10^{44}$ erg s$^{-1}$ is obtained from the unabsorbed best-fit model, $\xi=25^{+35}_{-11}$ erg cm s$^{-1}$ is the best-fit value (see Table~\ref{tab:bestfit}). Combining Eq.\ref{eq:dist} and Eq.~\ref{eq:densy} we obtain an upper limit for the density of the obscuring clump $n=(2.7\pm0.5)\times10^6$ cm$^{-3}$, which means that it is located at least at $r=(1.5\pm0.5)\times10^{18}\;\text{cm}=(0.50\pm0.15)\;\text{pc}=(8\pm2)\times10^3r_s$, where $r_s=2GM/c^2$ is the Schwarzschild radius. The estimated minimum size of the clump is therefore $R\simeq N_\text{H}/n=(6\pm2)\times10^{-3}$ pc.\\
\indent An estimate of the maximum distance of the cloud from the central source might be given by noting that the size of the clump cannot be larger than its distance from the X-ray source. Therefore we assume $N_\text{H}=nR<nr_\text{max}$ \citep[e.g.,][]{crenshaw12,tombesi13}. Substituting into the definition of ionization parameter we obtain
\begin{equation}
r_\text{max}=\frac{L_\text{ion}}{N_\text{H}\xi}.
\label{eq:rmax}
\end{equation}
\noindent Assuming again $L_\text{ion}\simeq1.5\times10^{44}$ erg s$^{-1}$, and the best-fit values of $N_\text{H}$ and $\xi$ (Table~\ref{tab:bestfit}), we obtain $r_\text{max}=30^{+40}_{-13}\;\text{pc}=5^{+5}_{-2}\times10^5\;r_s$.\\
\indent The increase of column density is likely due to a new clump of absorbing material, either due to a superposition of the WA and the E-UFO observed in Paper I, with an increased column density $N_\text{H}$, or an additional absorber located between the central source and the previously known absorber. The values of the minimum and maximum distance of the clump found in Eqs.~\ref{eq:dist} and \ref{eq:rmax} strongly suggest that the absorbing clumps is located outside the typical boundaries of the broad line region for a quasar of this luminosity. Indeed, we estimate a broad line region radius of $R_\text{BLR}\simeq0.07$ pc, considering the luminosity $\log (L_{5100\AA}/\text{erg s}^{-1})\simeq44.77$ of PG 1114+445 \citep{shen11}, once the relation between $R_\text{BLR}$ and $L_{5100\AA}$ derived by \cite{bentz09} is assumed. A comparison between the parameters of the absorber here detected and the ones measured in Paper I is shown in Fig.~\ref{fig:sum}.\\
\indent While no strong flux variability is detected between the {\it Swift}-XRT observations, minor variations are present (Figs.~\ref{fig:cr} and \ref{fig:lcfull}) and may be tied to the observed increased clumpiness. For instance, chaotic cold accretion (CCA) predicts fractal variations with power spectral density proportional to $f^{-1}$, where $f$ is the time frequency. In other words, smaller clumps are expected to contribute to the micro variations observed in the light curves \citep[e.g.,][]{gaspari17b}. The distance estimate puts the cloud at the meso scale in the CCA self-regulation framework \citep{gaspari20}. This is the typical transition scale at which CCA clouds become more clustered and collide frequently, thus generating flickering absorbers along the line of sight, starting from the X-ray/hot phase and potentially down to the radio/molecular phase \citep[e.g.,][]{tremblay18,rose19}.\\
\indent In the last decade many obscuring variable absorbers were found in the X-ray spectra of nearby AGN \citep[e.g.,][]{markowitz14}, with timescales ranging from decades \citep[e.g.,][]{kaastra14} to months or weeks \citep[e.g.,][]{matzeu16,mehdipour17,middei20} and even days \citep[e.g,][]{braito14,severgnini15}. These winds, with velocities than can be higher than typical WA velocities, are important because they may carry sufficient kinetic power to contribute to possible AGN feedback on the host galaxy. It is therefore important to continue monitoring these sources and possibly several others with current facilities such as {\it Swift} or eROSITA \citep{merloni12}, and with future missions such as the enhanced X-ray Timing and Polarimetry mission \citep[eXTP,][]{zhang19}, XRISM/Xtend \citep[e.g.,][]{yoneyama20} or the Athena Wide Field Imager \citep[WFI,][]{meidinger15} in order to look for both long-term and transient obscuration from ionized clumps. In the case of PG 1114+445, given the long-term nature of its obscuration, a monthly monitoring could help to identify lowest and highest states,  to study them with current X-ray telescopes such as {\it XMM-Newton}. In the future, forthcoming microcalorimeters such as Resolve on board XRISM \citep{xrism20} and Athena/X-IFU \citep{barret16} will be able to measure the X-ray spectrum with unprecedented energy resolution, letting us measure the outflow velocity of these obscurers with much higher accuracy, allowing to observe the full range of absorbers in AGN.

\begin{acknowledgements}
      We thank the referee for improving the paper with useful comments. We acknowledge financial contribution from the agreement ASI-INAF n.2017-14-H.0. EP acknowledges support from PRIN MIUR project "Black Hole winds and the Baryon Life Cycle of Galaxies: the stone-guest at the galaxy evolution supper", contract n. 2017PH3WAT. MG acknowledges partial support by NASA Chandra GO8-19104X/GO9-20114X and HST GO-15890.020-A grants. AT acknowledges the financial  support from FONDECYT Postdoctorado for the project n. 3190213. We acknowledge the use of public data from the {\it Swift} data archive. This research has made use of data and software provided by the High Energy Astrophysics Science Archive Research Center (HEASARC), which is a service of the Astrophysics Science Division at NASA/GSFC and the High Energy Astrophysics Division of the Smithsonian Astrophysical Observatory.
\end{acknowledgements}

%
   \bibliographystyle{aa} 
   \bibliography{biblio} 

\end{document}